%% file: python_binding.tex
\documentclass[sigconf,10pt]{acmart}
\setcopyright{none}
\copyrightyear{2021}
\acmYear{2021}
\acmConference[EuroMPI'21]{EuroMPI'21: 28th European MPI Users' Group Meeting}{September 7--8, 2021}{Virtual}
\usepackage{listings}
\usepackage{graphicx}

\begin{document}
\title{Generating Bindings in MPICH}
\author{Hui Zhou}
\affiliation{
  \institution{Argonne National Laboratory}
  \country{USA}
}
\email{zhouh@anl.gov}
\author{Ken Raffenetti}
\affiliation{
  \institution{Argonne National Laboratory}
  \country{USA}
}
\email{raffenetti@mcs.anl.gov}
\author{Wesley Bland}
\affiliation{
  \institution{Intel Corporation}
  \country{USA}
}
\email{wesley.bland@intel.com}
\author{Yanfei Guo}
\affiliation{
  \institution{Argonne National Laboratory}
  \country{USA}
}
\email{yguo@anl.gov}

\input{text/abstract}


\maketitle

\pagestyle{plain}

\input{text/intro}
\input{text/binding_toolbox}

\input{text/f08}
\input{text/qmpi}
\input{text/discussion}
\input{text/conclusion}

\section{Acknowlegment}
This material was based upon work supported by the U.S. Department of Energy, Office of
Science, under Contract DE-AC02-06CH11357.
% Intel probably want to put something here

\bibliographystyle{ACM-Reference-Format}
\bibliography{references}

\end{document}

%% file: text/abstract.tex
\begin{abstract}
The MPI Forum has recently adopted a Python scripting engine for generating the
API text in the standard document.  As a by-product, it made available
reliable and rich descriptions of all MPI functions that are suited for
scripting tools.  Using these extracted API information, we developed a
Python code generation toolbox to generate the language binding layers in
MPICH.  The toolbox replaces nearly 70,000 lines of manually maintained 
C and Fortran 2008 binding code with around 5,000 lines of Python scripts
plus some simple configuration.  In addition to completely eliminating code
duplication in the binding layer and avoiding bugs from manual code copying
, the code generation also minimizes the effort for API extension and code
instrumentation.  This is demonstrated in our implementation of MPI-4 large
count functions and the prototyping of a next generation MPI profiling
interface, QMPI.
\end{abstract}

%% file: text/intro.tex
\section{Introduction}
\label{sec:intro}
Code duplication is widely accepted as one of the top signs of bad software practice.  It inflates
the complexity and exponetially increases the burden of software maintainence.  However, in
practice, the ability to avoid code duplication is limited by the facility provided by the
programming languages.  Even in the case when the code can be refactored within the language, the
refactoring may add additional complexity that won't necessarily compensate the benefit of code
de-duplication.

% expand above sentence and lead into MPICH
One example is in MPICH\cite{website:mpich}'s binding layer. The binding layer refers to the code from the
MPI API function, e.g. \verb'MPI_Send', where application code calls upon, to where it calls into
the MPICH's internal C function, e.g. \verb'MPID_Send'.  This binding layer handles tasks such as
MPI profiling interface, parameter validation, MPI object handle conversion, and MPI error
behavior.  The code patterns among the 455 functions (MPI 3.1) are quite similar but often contains
sufficient variations that prevents us from replacing the duplicated patterns with macros or
functions.  These variations include trivial ones, e.g. all functions varies in its names
and argument prototypes, resulting in duplication in code blocks that deal with profiling interfaces
and error behavior. It also include non-trivial ones, e.g. the parameter validation logic,
which may differ from function to function even with the same parameter names and types.  As a
result, MPICH has been manually maintaining around 45,000 lines of such highly redundant code for
the C binding layer alone. 

% Need expand on what's going on with MPI Forum
% Explain that MPI Forum had the same 
A similar situation is encountered by the MPI Forum. The MPI Forum is an open group , with representatives from many organizations which defines and maintains the MPI standard.
The Forum maintains the MPI standard document in LaTeX source, inside which, it specifies all MPI functions in several programming languages. As the standard grows, the amount of API function increased significantly, the task of maintaining these function interfaces, which contains much redundancy, has also grown into a difficult and error-prone process. In MPI 4.0, it expanded the API by more than 100 large count functions that are very similar to the non-large functions but with varying details require effort to maintain. To make the task manageable, the MPI Forum adopted to develop a Python engine to generate document API specification.  Of particular importance, the Forum classified all MPI function parameters into a set of semantic oriented kinds, that are much richer than the language types, and it is readily accessible to programming tools. 

An example of this new semantic information is shown as in the following list.

{ \scriptsize
\begin{verbatim}
\begin{mpi-binding}
    function_name("MPI_Send")
    
    parameter("buf", "BUFFER",
        desc="initial address of send buffer",
        constant=True
    )
    parameter("count", "XFER_NUM_ELEM_NNI_SMALL",
        desc="number of elements in send buffer",
    )
    parameter("datatype", "DATATYPE",
        desc="datatype of each send buffer element"
    )
    parameter("dest", "RANK",
        desc="rank of destination"
    )
    parameter("tag", "TAG", desc="message tag")
    parameter("comm", "COMMUNICATOR")
\end{mpi-binding}
\end{verbatim}
}

Even though this is direct Python code, it is semantically clean and can be
easily parsed by a script. Most prominantly, now we have a semantic description
of each function parameters that is directly sourced from the MPI standard.

With the availability of the semantic database, we reevaluated the feasibility
of using scripts to generate MPICH's C binding layer. The new semantic
information greatly simplified the generation task. Instead of having to hard
code the specific logics for each functions, we can simply define the behavior
for each semantic kind, and generate the code accordingly. Based on this
assessment, MPICH team implemented a binding generation toolbox that generates
the C binding code and replaced around 45,000 lines of otherwise manually
maintained C code.

In addition to C binding generation, we also extended the toolbox to generate
MPICH's Fortran \verb'use mpi_f08' binding, which replaces additional 25,000 or
so lines of code that are used to be manually maintained. And with the binding
being generated, it is relatively straight forward to add support to the newly
introduced (in MPI 4) large count functions and the QMPI prototype. We'll
describe our design and implementations in dettail in the next section.
Finally, we'll discuss what we have learned in this effort and provide some
feedbacks to MPI Forum's future development in its python framework.

%% file: text/binding_toolbox.tex
\section{MPICH binding toolbox}
\label{sec:toolbox}
Traditionally in MPICH, the binding generations are limited to simpler bindings
such as Fortran binding and C++ binding. The generation scripts takes
information from mpi.h, which inlcudes all MPI function prototypes in C. The
typical tasks in these language bindings involve parameter type conversions,
which can be largely defined by the C types. However, variations still exist
between functions, and traditionally these variations are addressed by
hard-coding the function-specific logic inside the generation scripts.

The C binding, however, involves more challenges. In particular, it need deal with the parameter
validation that are not unique based on C types. For example, both MPI tags and MPI ranks are
integers, but they will require different validations. Without the semantic database, it will
require much more code logics to define individual variations among functions, making the code
generation too complex for its benefit.

As an initial effort, we aim to implement a new binding toolbox in
Python and generate the code in C binding and F08 binding, replace the
70,000 or so lines that are otherwise being manually maintained. The
first goal is to preserve the functionality, followed by a second goal
of implementing MPI 4.0 large count API and QMPI prototype. It is in the
plan that we eventually will replace the legacy Fortran binding scripts and use the
Python binding toolbox for the generation of Fortran \verb'mpif.h' and
\verb'use mpi' bindings as well.

\subsection{Design}
The design of our binding toolbox is illustrated in Figure.
The \verb'mpi-standard' git repository is not generally available to the
public, so it is not practical to either embed the latex sources or
import the repoistory as submodule. We choose to use a separate script
that parses the standard into two standard api configuration files to be
tracked within the MPICH repository. This transcribed version of APIs
are easier to work with than either the upstream LaTeX sources[reference] or
the Python scripts used by the MPI Forum, which is specialized for
generating LaTeX sources. Examples of the transcribed API text are
listed below.

{ \scriptsize
\begin{verbatim}
# mpi_standard_api.txt
...
MPI_Send:
    buf: BUFFER, constant=True,
        [initial address of send buffer]
    count: POLYXFER_NUM_ELEM_NNI,
        [number of elements in send buffer]
    datatype: DATATYPE,
        [datatype of each send buffer element]
    dest: RANK, [rank of destination]
    tag: TAG, [message tag]
    comm: COMMUNICATOR
...    
\end{verbatim}
}

If we reference the original Python code included in the LaTeX source,
this is nearly idential transcription but with Python syntax removed,
thus become language-neutral.

Each parameter is designated with a semantic ``kind'', e.g.
\verb'BUFFER'. The same C type, such as \verb'int' may be mapped from
different ``kind'', thus providing more semantic categoraizations.

The mapping from semantic ``kind'' to C types are extracted from the
upstream Python scripts and transcribed in \verb'apis_mapping.txt'.

{ \scriptsize
\begin{verbatim}
    # apis_mapping.txt
    LIS_KIND_MAPPING:
        BUFFER: choice
        ...
    BASE_C_KIND_MAP:
        BUFFER: void
        ...
    SMALL_C_KIND_MAP:
        .base: BASE_C_KIND_MAP
        POLYXFER_NUM_ELEM: int
        ...
    BIG_C_KIND_MAP:
        .base: BASE_C_KIND_MAP
        POLYXFER_NUM_ELEM: MPI_Count
        ...
    BASE_F90_KIND_MAP:
        ...
    SMALL_F90_KIND_MAP:
        ...
    BIG_F90_KIND_MAP:
        ...
    BASE_F08_KIND_MAP:
        ...
    SMALL_F08_KIND_MAP:
        ...
    BIG_F08_KIND_MAP:
        ...
\end{verbatim}
}

The mapping maps the ``kind'' to language-specific types. In particular,
the ``POLY-'' kind maps to different types depend on wether the function
is a large count variant.

Note that some information, e.g. whether a parameter should be a C
pointer is missing from the mapping table. The logic is embedded in the
Python code used by MPI Forum. This is due to the non-trivial difference
between how Fortran and C passes function parameters.
We ended up reproducing the same logic in our code. Ideally, we would
like an implementation-independent way of specifying this information.
Current decision is made to leave the transcribed APIs as close to
up-stream source as possible.

The APIs transcribed from mpi-standard are sufficient to generate the
standard document. As an implementation, MPICH will need some
customizattions. For example, not all functions should be generated, or
we may need generate additional \verb'MPIX_' extensions. We would also
like to add more ``directives'' to guide the generations to steer the
binding behavior. Examples include custom error checking, custom man
page notes, etc.
We want to avoid hard-coding these customizations inside Python scripts.
Therefore, we use another set of MPICH custom configurations to guide
the binding generation. The custom configurations files uses the same
format as the standard trascribed APIs, and are loaded by the same
routines from the toolbox. The entries are complimentary. The
customization can either add extra attributes, or override standard
attributes, or add new functions.
An example of custom file are listed below.

{ \scriptsize
\begin{verbatim}
# c/pt2pt_api.txt
MPI_Send:
    .desc: Performs a blocking send
    .seealso: MPI_Isend, MPI_Bsend
    .earlyreturn: pt2pt_proc_null
    ...
\end{verbatim}
}

Only the entries listed in one of the custom files will be generated.
This is particularly helpful when we implement a new API version, e.g.
MPI 4, not all new functions are implemented yet. The separate
customization files gives us some logical organizations, which is
desirable for code maintainence.  There are quite a few customization
semantics that are designed to balance between manual maintainence and
increase of script complexities. We'll cover some of them in more
details in the next sections.

In a overview, the binding toolbox loads standard APIs and custom APIs,
then generate the binding source files. It can be either invoked during
``autogen'' -- a step run by developers before distributing the
releases, or run during ``configure''. The latter allows taking options
that are customed to user's build environment.

\includegraphics[width=7.5cm]{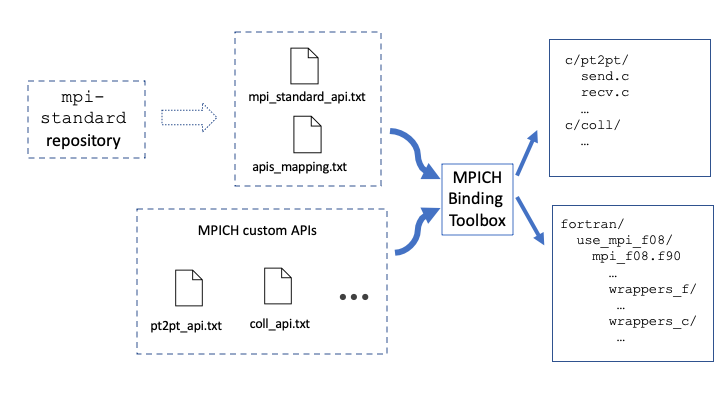}

\subsection{C binding}
The term ``C binding'' is not quite precise. For other languages, e.g.
Fortran binding, refers to the code that starts as in one language,
internally calls the C API functions, and returns the results back to
the original language.
MPICH is a C library. Why do we need a C binding?
This is because there are some ``standard'' tasks to be done for nearly
every MPI functions. The tasks here refers to a general sense that it
requires some ``code'', but not necessarily mean a runtime operation.
For example, one of the ``task'' is to handle MPI profiling interface.
Depend on build environment, we may need declare weak symbols, or define
macros to rename the function name. The ``task'' is just compile time
manipulation, but requires some code maintainence nevertheless.

An overview of the C binding is listed as following.

{\scriptsize
\begin{verbatim}
/* -- send.c -- */
/* -- Profiling Block -- */
#if defined (HAVE_PRAGMA_WEAK)
#prama weak MPI_Send = PMPI_Send
#elif ...
...

/* -- Man Page Documentation -- */
/*@
   MPI_Send - Performs a blocking send
...
 */

/* -- function definition -- */
int MPI_Send(...) {
    [local variables]
    [logging]
    [enter critical section]
#ifdef HAVE_ERROR_CHECKING    
    [validate each parameters]
#endif
    [convert handle objects to internal pointers]

    [handle trivial cases, e.g. dest is MPI_PROC_NULL]

    /* ... body of routines ... */
    mpi_errno = MPID_Send(...);

    [handle errors]
}
\end{verbatim}
}

Ideally, we would like the scripts to automatically generate all these
parts and the custom configuration can be reduced to

{ \scriptsize
\begin{verbatim}
MPI_Send:
    .desc: Performs a blocking send
\end{verbatim}
}

Here, an attribute that starts with ``.'' denotes an attribute for the
function. This is necessary to differentiate from a parameter.
For many standard behaviors, we can simply add custom attributes. For
example, \verb'.seealso: ...' adds the seealso links to the man page. As
another example, \verb'.earlyreturn: pt2pt_proc_null' adds a standard
check above the ``body of routine'' to check for \verb'MPI_PROC_NULL'
and immediately return if it's the case. The script is able to figure
out the simple logic such as which parameter to check.

However, the customization needs to be more verbose. For example, we
often want to add specific man page notes and custom body of routines.
Neither can fit into a simple attribute.
There are a few design goals.
First, we would like to keep the config file format simple, for the
least, it is not desirable to add programming syntax into config files
result in writing code in awkward constructs.
Second, we want the custom documentation notes maintained as notes and
custom code block maintained as code. Introducing extra mark-ups will
defeat our purpose of ease the project maintainence.
The solution is to introduce comment blocks and code blocks in the
config files, as in the following listing.

{ \scriptsize
\begin{verbatim}
MPI_Send:
    .desc: Performs a blocking send
/* -- notes-1 --
    Notes:
    ...
*/
{ -- body_of_routine --
    ...
}
\end{verbatim}
}

The syntax marked by ``/* ... */'' and ``\{ ... \}'' directly matches
the syntax we used in C source file. The annotation marked by `--` can
be used to designate where the custom comment or code block should be
placed.

[highlight the complexity in parameter validation, especially in collectives]

% large count in here
\subsection{Implementing large count API}
One of trigger for this work is the introduction of large count
functions in MPI 4.0. For each MPI function that contains a count or
displacement parameter, the MPI 4.0 standard provide a companion large
count version. For example, \verb'MPI_Send_c' is provided to allow
directly passing a large \verb'count' argument into the function, which
otherwise won't fit into the parameter type in \verb'MPI_Send', which
uses C `int`.
This introduces more than 100 new functions into the MPI standard.

The prospect of manually adding these functions is daunting. It is
further complicated by the fact that it is undesirable to use
\verb'MPI_Count' internally since \verb'MPI_Count' potentially can be
wider than system's address-sized integer, and it is both impractical
and may degrade the library performance. For this reason, we internally
use \verb'MPI_Aint' for most of the count-type arguments. Thus, the
binding layer need take care of the potential integer type conversions,
which depends on the actual type of either \verb'MPI_COUNT' or
\verb'MPI_Aint'. We were looking at the potential explosion of
boiler-plate code.

With the new binding toolbox, generation of all the large count
functions are straight forward. The ``kind'' attribute can tell whether
a function will require a large count version, and we need to call the
generation routines twice -- once for regular function, and once for
large count function. The additional code in the script is still
non-trivial due to the conversion requirement described above. However,
we are able to factor the new logic cleanly from the rest of the
complications.

% delete the mpix section, disperse the points somewhere
\subsection{MPIX namespace}
The binding toolbox also makes adding experimental extensions simple.
Other than its experimental nature, \verb'MPIX' functions are otherwise
treated the same as regular \verb'MPI' functions, require proper
declaration and visibility, require the MPI error behavior, and require
to be available to the Fortran bindings. With the binding toolbox,
adding a \verb'MPIX' function in most case is simply add one or two
lines in the custom config file provided the internal function is
already implemented.   

A recent twist further demonstrated the convenience of having the
binding generated. We released MPICH 4.0a1 in anticipation of official
ratification of MPI 4.0 specification. However, the official
ratification was delayed in the last week. By convention, that means we
need release MPICH with MPI 4.0 functions as MPIX functions. We need
rename the headers, bindings, and the corresponding tests. There are
many new functions introduced by MPI 4.0 -- 39 new functions and more
than 100 new large count functions. With the binding toolbox, we are
able to simply prepare a table of these functions that need MPIX
prefixes, and have the script generate the binding correspondingly.

%% file: text/f08.tex
\section{Fortran 2008 binding}
MPICH provides three separate Fortran language bindings. One via \verb'include mpif.h',
one via \verb'use mpi', and one via \verb'use mpi_f08'. Each binding
progressively supports more modern Fortran features. The first one often
referred as F77 binding, the second one as F90 binding, and the last one as F08
binding. These references can be misleading, however. In particular,
\verb'include mpif.h' interface can be used directly in modern Fortran code, and
some requirement, e.g. both the \verb'include' feature and C linking
conventions, are not part of F77 standard. Here we'll refer to these bindings as
the \verb'mpif.h' binding, the verb'use mpi' binding, and \verb'use mpi_f08'
binding.

Other than defining the MPI
constants, both the \verb'mpif.h' binding and \verb'use mpi' binding are
based on non-standard C/Fortran linking conventions. The bindings defines a set
of C interface functions that directly links to the Fortran applications. Both
bindings are traditionally generated by legacy scripts in MPICH based on C
prototypes in \verb'mpi.h'. Because the Fortran binding only need to take care
of the parameter conversion based on Fortran/C langauge conventions, it is
less complex. Some parameters may take on special MPI constants and has to be
separately checked and replaced by the corresponding C constants. These
parameter semantics are not provided by \verb'mpi.h' and are conventionally
manually listed in the generation script. We anticipate the legacy scripts can
be greatly simplified now that the semantic \verb'kind' information is
available. 

The \verb'use mpi_f08' binding utilizes the modern Fortran features, in
particular, the standard C interoperability feature since Fortran 2003 and the
techinical specification TS 29113\cite{TS29113} that adds support for choice
buffers interoperability. The interoperability feature allows the binding to
directly specify the C interface in Fortran and the compiler will provide type
checking, thus greatly improves type safety in an MPI Fortran application.
The implementation of MPICH's Fortran 2008 binding has been discussed in detail
in \cite{Zhang-F08binding}.  Here we just give an overview. The F08 binding
declares C routines in a interface block using Fortran's C interoproperability
feature. It allows directly passing most of the Fortran argument directly into
C functions with the necessary type safety provided by the language. It is
still necessary to have wrapper functions in Fortran to do necessary type
conversions, and sometimes, checking and replacing MPI constants. For MPI
functions with choice buffers, additional C wrapper routines are defined to
query information on the Fortran assumed-type, assumed rank arguments.

The generation of F08 binding is more demanding than the other Fortran bindings.
Both the Fortran interface and the C interface need to be fully specified. In
addition, the parameter in/out directions, the choice buffer semantics, special named constant values, and
assumed array lengths are needed to correctly generate the correct binding
interface and they can't be derived from C prototypes from \verb'mpi.h'. Thus,
most F08 binding code in MPICH was manually maintained rather than generated.

Fortunately, all these extra informations are now provided by the MPI standard
python layer. Extending the MPICH binding generation toolbox to generate F08
binding is straight forward. With less than 2000 lines of additional Python
scripts, we are able to replace more than 25,000 lines of manual code.
In addition to maintain the same features described in detail in
\cite{Zhang-F08binding}, we are able to easily add the missing functions that
previously missed due to the manual process. It is also simple to add the large
count functions via Fortran generic interface.

%% file: text/qmpi.tex
\section{Prototyping QMPI}
The main goal of having bindings generated by scripts is to avoid code
duplication and accomodate orthoganalities between features. This is
demonstrated by the implementation of QMPI\cite{QMPI} prototype. QMPI is a proposed next
generation MPI profiling interface currently under development by MPI Forum Tools
working group.  Similar to the current PMPI interface, QMPI allows
tools to intercept MPI function calls at runtime. Unlike PMPI, QMPI
allows for simutaneous attachment of multiple tools, which removes a
key limitation of PMPI.  Until QMPI officially replaces PMPI, we need
to ensure the prototype QMPI implementation works alongside the
existing PMPI interface in MPICH. This poses a challenge since the
PMPI implemention differs based on compiler and operating system
features, resulting in special PMPI code in hundreds of files. For
instance, PMPI functions can be declared as weak symbols, in which
case there is only a single function definition for each API. Or, PMPI
functions can be an actual separate functions, defined using
preprocessor directives. In order to add QMPI support, we needed a
flexible scheme that worked for all situations. % does the PMPI implementation affect QMPI in any way?
In fact, we experimented with a few different schemes during
development. This would not have been possible to do efficiently if
the C binding was still scattered as hundreds of separate C
files. Trying a scheme in that case would have meant modifying each of
these files separately numerous times.  With the new binding toolbox,
the logic of profiling schemes are orthoganol to other complexities
such as error checking and parameter conversions. Changes could be
made independently and without concern for impacting unrelated
portions of the code, allowing for a more rapid development and
feedback cycle.

%% file: text/discussion.tex
\section{Discussion}
\label{sec:discussion}
The binding generation toolbox is successful in its goal of removing
redudant code and refactor complexities into orthogonal components.
However, at nearly 4,500 lines of Python scripts, it is still more
complex than desired.  This is especially true to developers new to
the project when they need to decipher the indirect logic from Python
scripts rather than the direct C code or Fortran code.
% probably move to future work
We believe the direction is to further evolve the custom API config
files into templates. The Python layer may need expansion to support the
extra templating features, but we will be able to extract the output
code logic from the Python scripts, thus can be better maintained.
We are testing this template idea by allowing direct comment blocks and
code blocks in the config files. What's missing is names and macro
parameters.

It is undesirable that we are basing our binding generation on a derived
API format rather than direct source from upstream.
This is mainly due the fact that MPI forum does not expose the API as
language neutral information, but rather as an internal Python code. It
should be noted that the MPI Forum's Python tool can optionally export the
data as a json file, and our binding toolbox is able directly load this
json file instead of the custom derived format. We still prefer to use
the derived format due to a few shortcoming of this json export. One,
it is very verbose. It is 56,253 lines long, compared to our
\verb'mpi_standard_api.txt' at 2,690 lines long. Two, the json format
is not easily readable or editable by developers. We believe this is
the same reason that MPI Forum prefers to embed direct Python code
rather than embedding the json format.

In addition, only the function API is exported in the json file. The
langauge type mapping, necessary for actual binding generation is
embedded in Forum's Python code and has to be separately extracted.

Separate from the format, much of the parameter validation logic is
missing from what is provided by the forum. For example, the
validation of \verb'RANK' depends on whether it is a ``send rank'',
which allow the value of \verb'MPI_PROC_NULL', or it is a ``recv
rank'', which in addition allows the value of \verb'MPI_ANY_SOURCE',
or a normal rank, which none of the special value is allowed. We
currently hard-code these nuisances checking both the parameter
``kind'' and function names. This hard-coded logic is fragile and
ideally should be separated out with configuration.

It is understandable since the main purpose of forum's adoption of
this API layer is to facilitate generation of \LaTeX document, rather
than generating MPI library code. We hope that we have demonstrated
that general availability of script friendly API information is very
useful to implementations and MPI researchers, and provide some
motivations for the Forum API layer to evolve for more general
purposes.

%% file: text/conclusion.tex
\section{Conclusion}
\label{sec:conclusion}
We have implemented a new binding generation toolbox in MPICH
utilizing the newly available script-friendly API information from the
MPI standard document source code. This allowed us to replace some 70,000
lines of otherwise manually maintained code with 5000 lines or so of
scripts. The binding generation removes large amounts of code
duplication, and brings more consistency to the MPICH project
overall. With a binding generation toolbox, we were able to easily
extend the language bindings to support new API functions from the
upcoming MPI 4.0 standard, and simplify the process of prototyping a
new tool interface -- QMPI.  While we recognize the overall success,
we also noted the current shortcoming of having to base our binding
generation on a derived source rather than directly from
upstream. Ultimately, It would be beneficial for the MPI standard to
expose more semantics in the API specification so the complexities of
our generation script can be further simplified.